\documentclass[12pt]{article}
\usepackage[dvips]{graphicx}

\begin{document}

%\titlerunning{Title running}

\begin{center}
{\Large \textbf{\boldmath KINETIC DESCRIPTION OF W AND Z BOSON VACUUM
CREATION IN THE EARLY UNIVERSE}} %<== title (bold face, capitalize)
\vskip 5mm

D.~B.~Blaschke{$^{1,2}$, M.~P.~D\c{a}browski$^{3}$, V.~V.~Dmitriev$^{4}$ and 
S.~A.~Smolyansky$^{4\dag }$ }\\[0pt]
{\small (1) \textit{Institute of Theoretical Physics, University of
Wroc{\l}aw, Wroc{\l}aw, Poland } \\[0pt]
(2) \textit{Bogoliubov Laboratory of Theoretical Physics, JINR, Dubna, Russia } \\[%
0pt]
(3) \textit{Institute of Physics, University of Szczecin, Szczecin, Poland}
\\[0pt]
(4) \textit{Physical Department of Saratov State University, Saratov, Russia
} \\[0pt]
$\dag $ \textit{E-mail: smol@sgu.ru }}
\end{center}

\vskip 5mm

\begin{center}
\begin{minipage}{150mm}
\centerline{\bf Abstract}
We consider an alternative mechanism for the production of the cosmic microwave background (CMB)
radiation. It is basically due to vacuum pair creation (VPC) of vector bosons (W and Z)
as a consequence of a rapid W and Z mass generation during the electroweak
phase transition in the early Universe.
The mechanism is as follows: after their pair crreation, the vector bosons may either annihilate
directly into photons or decay into leptons and quarks which subsequently annihilate as lepton-antilepton
and quark-antiquark pairs into photons.
Preliminary estimates show that the number of CMB photons obtained this way can be
sufficient to explain the presently observed CMB photon density.
In this contribution we present an exactly soluble model for vacuum pair creation kinetics.
%Details depend on a fine adjustment of the  matter background equation of
%state (EoS), some features of the electroweak phase transition, and the
%initial values of the cosmological evolution.
\end{minipage}
\end{center}

\vskip 10mm

The mechanism of CMB\ generation by means of primordial massive vector (W
and Z) boson pair creation with subsequent decay and annihilation processes
was proposed in Refs. \cite{1,2}. The motivation for such a scenario is that
these particles are heavier than other electrically charged particles and
therefore they are born earlier in the process of cosmological expansion.
Preliminary estimates performed in \cite{2}, within the framework of a conformal
cosmological model, show on a qualitative level that such a mechanism of
CMB\ generation can indeed be effective.

In the present work it is shown that this conclusion remains valid in the
framework of the standard cosmological model in the FRW\ space. We use the
system of kinetic equations for the description of the VPC of massive vector
bosons in a nonstationary metric \cite{birell,3,4} as a methodical basis. The
nonstationary gravitational field itself provides a source for vacuum pair
creation.
However, there is an additional mechanism of VPC based on the time
dependence of the particle rest masses \cite{5,6,7} during the electroweak phase
transition.
Since at present many details of the evolution of the Universe during the
electroweak phase transition remain unknown, we use a Heaviside step function
type time dependence for the rest mass of W and Z bosons (their mass difference is
neglected here).
On the other hand, the description of the longitudinal component of the massive
vector field within the Wentzel model in the FRW metric gives considerable
difficulties (all observable values related to particles with
longitudinal polarization are described by divergent integrals).
For the present discussion we assume that this problem may be solved and that
the contribution of the longitudinal component is of the order of the
transversal ones, as in the conformal model \cite{2,Blaschke:2004ia,dabr}.
In this work we show that under these assumptions together with reasonable
parameter values characterizing the time dependence of the scale factor and
the mass, the inertial mechanism can indeed provide qualitative agreement
with the observed CMB\ photon density.
Preliminary results were reported in Ref. \cite{7.1}.

The VPC of massive vector bosons is described by the kinetic equation \cite{3,4}
\begin{equation}
f^{\prime }(J,\eta )=\frac{1}{2}w(J,\eta )\int_{\eta _{0}}^{\eta }d\eta
^{\prime }w(J,\eta ^{\prime })\left[ 1+2f(J,\eta ^{\prime })\right] \cos
2\theta (J;\eta ,\eta \prime )~,  \label{KE}
\end{equation}%
where
\begin{equation}
w(J,\eta )=\frac{\omega ^{\prime }(J,\eta )}{\omega (J,\eta )}=\frac{%
m^{2}(\eta )a^{2}(\eta )}{\omega ^{2}(J,\eta )}\left[ \frac{a^{\prime }(\eta
)}{a(\eta )}+\frac{m^{\prime }(\eta )}{m(\eta )}\right] ~,  \label{w}
\end{equation}%
\bigskip $a(\eta )$ and $m(\eta )$ are the scale factor and rest mass
depending on the conformal time $\eta $,
\begin{equation}
\omega (J,\eta )=\left[ \lambda ^{2}+m^{2}(\eta )a^{2}(\eta )\right] ^{1/2}~.
\label{omega}
\end{equation}%
The distribution function $f\left( J,\eta \right) $ is
defined in the corresponding phase space of quantum numbers
$J=\left\{\lambda ,l,m\right\} $, where $\lambda $ is the momentum
quantum number that corresponds to the physical momentum $k=\lambda /a(\eta )$.
The last term in Eq. (\ref{w}) is due to the inertial mechanism of vacuum
particle creation \cite{7}.
The remaining notations are given in Refs. \cite{3,4}.
We omitted the polarization indices in Eqs. (\ref{KE}), (\ref{w}) according to
the hypothesis about the equality of longitudinal and transversal components.

Since all fields were massless right up to the electroweak phase transition
\cite{8}, we suggest the equation of state of radiation $p=\varepsilon /3$ to be
valid at that stage of the evolution. The corresponding scale factor in such a case is
\begin{equation}
a(\eta )=a_{1}\sinh (\eta ),\quad t=a_{1}(\cosh (\eta )-1).  \label{a}
\end{equation}%
We assume that the rest mass of the vector bosons is changed according to a Heaviside
step function law
\begin{equation}
m(\eta )=m_{f}\ \Theta (\eta -\eta _{\mathrm{EW}}),  \label{m}
\end{equation}%
as a consequence of electroweak phase transition, which occurs
"instantaneously" at the time $t_{\mathrm{EW}}=10^{-10}$ s. It is assumed
that vector bosons are massless up to the phase transition and their
final masses are equal to the presently observable ones, $m_{f}=m_{W}$.
According to \cite{8.1}, we take $\Theta (0)=1$ at $\eta =\eta _{\mathrm{EW}}$, which
corresponds to an ordinary definition of the quasiparticle energy
(\ref{omega}) at $\eta <\eta _{\mathrm{EW}}(\omega (J,\eta )=\lambda )$
and at $\eta >\eta _{\mathrm{EW}}$ (see Eq. (\ref{wew}) below).

The kinetic equation (\ref{KE}) can be solved exactly in this model, resulting in
\begin{equation}
f\left( J,\eta \right) =\frac{1}{2}\frac{\Theta (\eta -\eta _{\mathrm{EW}})}{%
\omega _{\mathrm{EW}}^{4}/m_{f}^{4}a_{\mathrm{EW}}^{4}-1}~,
\label{f}
\end{equation}
where
\begin{equation}
\omega_{\mathrm{EW}}=\sqrt{\lambda ^{2}+m_{f}^{2}a_{\mathrm{EW}}^{2}}~,\quad
a_{\mathrm{EW}}=a(\eta _{\mathrm{EW}})~.
\label{wew}
\end{equation}
In a vicinity of the phase transition, $\eta \approx \eta _{\mathrm{EW}}$,
the first term on the r.h.s. of Eq.~(\ref{w}) being proportional
to $a^{\prime }/a$ can be neglected in comparison to the second one.

We are interested in the number density of the primordial W\ and Z\ bosons
\cite{3,4}~,
\begin{equation}
n\left( \eta \right) =\frac{g}{2\pi ^{2}a^{3}\left( \eta \right) }\int d\mu
\left( \lambda \right) f\left( J,\eta \right) ~,  \label{n}
\end{equation}
where $g=3\cdot 3=9$, is the boson degeneracy factor (three isospin and three
polarization degrees of freedom).
The measure $d\mu \left( \lambda \right) $ in (\ref{n}) depends on a cosmological model.
In the quasi-Euclidean space we have $d\mu =d\lambda \ \lambda ^{2}$
and the substitution of Eq.~(\ref{f}) into Eq.~(\ref{n}) leads to the result
\begin{equation}
n\left( \eta \right) =\frac{g}{4\pi ^{2}a^{3}\left( \eta \right) }
\int\limits_{0}^{\infty }\frac{d\lambda \ \lambda ^{2}}{(\lambda
^{2}/m_{f}^{2}a_{\mathrm{EW}}^{2}+1)^{2}-1}=\frac{9\,}{\sqrt{2}\pi }\left(
\frac{a_{\mathrm{EW}}m_{f}}{2a(\eta )}\right) ^{3}.  \label{n2}
\end{equation}
This result is an upper estimate for the total particle number density
because the increase in the vector meson mass during the electroweak phase
transition is "smoother" than a Heaviside step function law says.
It is assumed that the density (\ref{n2}) is equal to that of the relic photons as a consequence of a complete
annihilation of the primordial vector boson pairs.

In the cosmological model presented here, the measure in the integral (\ref{n})
is $d\mu =d\lambda \ (\lambda ^{2}+1)$, which leads to an infrared
divergence of this integral.

It is further assumed, that the particle number density for later times is
changed as a result of the expansion of the Universe, i.e. the present day
photon number density equals $n_{0}=n_{\mathrm{EW}}\left( a_{\mathrm{EW}%
}/a_{0}\right) ^{3}$, where $a_{0}$ is the present scale factor of the
Universe. According to Eq.~(\ref{n2}) we obtain $n_{\mathrm{EW}}=n(\eta _{%
\mathrm{EW}})\sim 10^{46}~\mathrm{{cm^{-3}}}$.

The important role of electroweak interactions in the early
Universe has already been emphasized in the literature (see e.g. \cite{12}).
The ratio $a_{_{\mathrm{EW}}}/a_{0}$ can easily be obtained through the recombination
point at $t_{\mathrm{R}}=10^{13}$ s, when the radiation-dominated era ($a\sim t^{1/2}$)
changes into the matter-dominated era ($a\sim t^{2/3}$), i.e.,
\begin{equation}
\frac{a_{\mathrm{EW}}}{a_{0}}=
\frac{a_{\mathrm{EW}}}{a_{\mathrm{R}}}\frac{a_{\mathrm{R}}}{a_{0}}
\sim \left( \frac{10^{-10}\ \mathrm{s}}{10^{13}\ \mathrm{s}}\right)^{1/2}
\left( \frac{10^{13}\ \mathrm{s}}{10^{18}\ \mathrm{s}}\right)^{2/3}.
\end{equation}
Taking into account that $a_{\mathrm{EW}}/a_{0}\sim 10^{-15}\div 10^{-14}$,
we obtain a range of densities $n_{0}\sim 10\div 10^{4}~\mathrm{{cm^{-3}}}$, which
includes the presently observed CMB photon density.

In conclusion, we have shown that the presently observed relic photon density can be
obtained from the inertial mechanism of particle production according to
Eqs.~(\ref{m}) and (\ref{a}). Note that within the above scenario the
massive vector bosons are created from the vacuum due to the inertial
mechanism only. In other words, the W and Z bosons are created indirectly
by the gravitational field in an empty Universe. Due to their
subsequent weak decay into quarks and leptons, the cosmic matter content
would arise with photons stemming from annihilation and decay processes.
While the details of the kinetics of such processes remains to be worked
out, there is another appealing aspect of this scenario: the triangle
anomaly for the vector bosons might provide a mechanism \cite{1} for the
generation of the baryon and lepton asymmetry in the early Universe.


\begin{thebibliography}{99}
\bibitem{1} %\bibitem{Blaschke:2004by}
D.~B.~Blaschke, S.~I.~Vinitsky, A.~A.~Gusev, V.~N.~Pervushin and
D.~V.~Proskurin,
%``Cosmological Production of Vector Bosons and Cosmic Microwave Background
%Radiation,''
Phys.\ Atom.\ Nucl.\ \textbf{67}, 1050 (2004)
%  [Yad.\ Fiz.\  {\bf 67}, 1074 (2004)]
[arXiv:hep-ph/0504225];
%%CITATION = YAFIA,67,1074;%%
\newline
D.~Blaschke, V.~Pervushin, D.~Proskurin, S.~Vinitsky and A.~Gusev,
%``Cosmological creation of vector bosons and fermions,''
arXiv:gr-qc/0103114.
%%CITATION = GR-QC/0103114;%%

%\bibitem{1} V.~N. Pervushin, Acta Physica Slovakia, \textbf{53}, 237 (2003).

%\cite{Blaschke:2005jf}

\bibitem{2} %\bibitem{Blaschke:2005jf}
D.~B.~Blaschke, A.~V.~Prozorkevich, A.~V.~Reichel and S.~A.~Smolyansky,
%``Kinetic description of vacuum creation of massive vector bosons,''
Phys.\ Atom.\ Nucl.\ \textbf{68}, 1046 (2005).
%  [Yad.\ Fiz.\  {\bf 68}, 1087 (2005)].
%%CITATION = YAFIA,68,1087;%%

\bibitem{birell} N.~D. Birrell and P.~C.~W. Davies, Quantum Fields in Curved Space,
Cambridge University Press, 1982.

\bibitem{3} A.~A. Grib, S.~G. Mamaev, M.~V. Mostepanenko. Vacuum Quantum
Effects in Strong External Fields. St. Petersburg: Friedman Lab. Publ. (1994).

\bibitem{4} A.~V. Veriaskin, V.~G. Lapchinsky, V.~A. Rubakov, Preprint
P-0198. Institut of Nuclear Research (1981).

%\cite{Behnke:2001nw}

\bibitem{5} %\bibitem{Behnke:2001nw}
D.~Behnke, D.~B.~Blaschke, V.~N.~Pervushin and D.~Proskurin,
%``Conformal cosmology and supernova data,''
Phys.\ Lett.\ B \textbf{530}, 20 (2002). %  [arXiv:gr-qc/0102039].
%%CITATION = PHLTA,B530,20;%%

%\cite{Barbashov:2005hu}

\bibitem{6} %\bibitem{Barbashov:2005hu}
B.~M.~Barbashov, V.~N.~Pervushin, A.~F.~Zakharov and V.~A.~Zinchuk,
%``Hamiltonian cosmological perturbation theory,''
Phys.\ Lett.\ B \textbf{633}, 458 (2006). %  [arXiv:hep-th/0501242].
%%CITATION = PHLTA,B633,458;%%

%\cite{Filatov:2007ha}

\bibitem{7} %\bibitem{Filatov:2007ha}
A.~V.~Filatov, A.~V.~Prozorkevich, S.~A.~Smolyansky and
V.~D.~Toneev,
 Phys.\ Part.\ Nucl. \textbf{39},  886 (2008).
%``Inertial mechanism: dynamical mass as a source of particle creation,''
%arXiv:0710.0233 [hep-ph].
%%CITATION = ARXIV:0710.0233;%%

%\bibitem{7} A.V. Filatov, A.V. Prozorkevich, S.A. Smolyansky, V.D. Toneev,
%Phys of Elem. Part. and Atom Nucl., in press.

%\cite{Blaschke:2004ia}

\bibitem{Blaschke:2004ia} D.~B.~Blaschke, A.~V.~Prozorkevich and
S.~A.~Smolyansky,
%``Vacuum creation of massive vector bosons and its application to a
%conformal cosmological model,''
arXiv:hep-ph/0411383. %%CITATION = HEP-PH/0411383;%%

\bibitem{dabr} M.~P. D\c{a}browski, J. Garecki and D.~B. Blaschke,
arXiv:0806.2683[gr-qc], Ann. Phys. (Berlin), in press (2008).
%"Conformal transformations and conformal invariance in gravitation"
%%CITATION = ARXIV:0806.2683;%%

\bibitem{7.1} D.~B. Blaschke, V.~V. Dmitriev, A.~V. Prozorkevich and S.~A.
Smolyansky, ''Kinetics of vacuum creation of W and Z bosons in early Universe'',
Talk given at the XIII International Conference Selected Problems of Modern
Theoretical Physics, Bogoliubov Laboratory of Theoretical Physics, Dubna,
Russia, June 23-27, (2008).

\bibitem{8} A.~Linde. Particle Physics and Inflationary Cosmology. Chur:
Harwood Academic Publishers (1990).

%\cite{Kibble:1980mv}

\bibitem{8.1} G. Doetsch, Anleitung zum praktischen Gebrauch der
Laplace-Transformation, M\"unchen: Springer (1967); G.B. Arfken and H.J. Weber,
Mathematical Methods for Physicists, Academic Press (2001).

%\bibitem{9} %\bibitem{Kibble:1980mv}
%T.~W.~B.~Kibble, %``Some Implications Of A Cosmological Phase Transition,''
%Phys.\ Rept.\ \textbf{67}, 183 (1980). %%CITATION = PRPLC,67,183;%%

\bibitem{12} A.R. Liddle, D.H. Lyth, Cosmological Inflation and Large Scale
Structure, Cambridge: Cambridge University Press (2000).
\end{thebibliography}
\end{document}